# DXL: a sounding rocket mission for the study of solar wind charge exchange and local hot bubble X-ray emission

M. Galeazzi, M. Chiao, M. R. Collier, T. Cravens, D. Koutroumpa, K. D. Kuntz, S. Lepri, D. McCammon, F. S. Porter, K. Prasai, I. Robertson, S. Snowden, Y. Uprety

**Abstract** The Diffuse X-rays from the Local galaxy (*DXL*) mission is an approved sounding rocket project with a first launch scheduled around December 2012. Its goal is to identify and separate the X-ray emission generated by solar wind charge exchange from that of the local hot bubble to improve our understanding of both. With 1,000 cm$^2$ proportional counters and grasp of about 10 cm$^2$ sr both in the ¼ and ¾ keV bands, DXL will achieve in a 5-minute flight what cannot be achieved by current and future X-ray satellites.

\-\-\-\-\-\-\-\-\-\-\-\-\-\-\-\-\-\-\-\-\-\-\-\-\-\-\-\-\-\-\-\-\-\-\-\-\-\-\-\-\-\-\-

*M. Galeazzi (✉), K. Prasai, Y. Uprety*
*University of Miami, Coral Gables, FL, USA*
*galeazzi@physics.miami.edu, tel: +1-305-284-2326, fax: +1-305-284-4222*

*M. Chiao, M. R. Collier, D. Koutroumpa, F. S. Porter, S. Snowden*
*NASA/Goddard Space Flight Center, Greenbelt, MD, USA*

*T. Cravens, I. Robertson*
*University of Kansas, Lawrence, KS, USA*

*K. D. Kuntz*
*Johns Hopkins University, Baltimore, MD, USA*

*S. Lepri*
*University of Michigan, Ann Arbor, MI, USA*

*D. McCammon*
*University of Wisconsin, Madison, WI, USA*

\-\-\-\-\-\-\-\-\-\-\-\-\-\-\-\-\-\-\-\-\-\-\-\-\-\-\-\-\-\-\-\-\-\-\-\-\-\-\-\-\-\-\-

# 1. Introduction

The first detection of extrasolar X-rays came in 1962 (Giacconi et al. 1962), showing the presence of a diffuse X-ray flux. Observation of diffuse emission was extended to energies below 1 keV identified by a series of three papers in 1968-69 (Boyer et al. 1968, Henry et al., 1968, Bunner et al. 1969), resulting in the discovery of Galactic emission. Subsequent investigations extensively studied the nature of the diffuse emission and its correlation with the neutral material in the Milky Way (McCammon & Sanders 1990, and references therein). X-rays observed in the ¼ keV band in the direction of the Galactic plane must originate locally as they are easily absorbed by the neutral material of the Galactic disk. It is currently believed that a significant fraction of the ¼ keV emission originates in an irregularly shaped cavity in the neutral hydrogen of the Galactic disk which contains the Sun, dubbed the Local Hot Bubble (LHB). Despite the "local" origin of the emission, there is still a significant uncertainty in its nature and characteristics. The reason for this uncertainty lies primarily in the nature of the emission, which is weak and diffuse, and its superimposition on emission from other sources of the Diffuse X-ray Background (DXB). In particular, it has been essentially impossible to isolate the LHB contribution from the emission from Solar Wind Charge eXchange (SWCX) recombination. SWCX emission is generated when the highly ionized solar wind interacts with neutral or near-neutral gas in the Earth's exo-atmosphere and in the interplanetary medium. An electron is captured from a neutral atom (H or He) into an outer orbital of the highly ionized solar wind species in a quasi-resonant process and cascades to the ground state of the ion, often emitting soft X-rays in the process. The SWCX spectrum is therefore dominated by characteristic line emission that is very difficult to separate from the presumably thermal emission of the LHB.

It is, however, possible to separate the foreground SWCX emission from the extra-solar sources, such as the LHB, by looking at its spatial and temporal signatures. The slowest time-varying, and thus most troublesome, component of the SWCX emission originates in the interplanetary medium and should show a significant geometric variation due to the focusing of interplanetary helium. Interstellar neutrals flow through the solar system due to the motion of the heliosphere through the local interstellar cloud. Gravity significantly affects helium trajectories which execute Keplarian orbits and form a "focusing cone" downstream of the Sun centered at ~$6^o$ below the ecliptic plane. This results in a localized downstream enhancement of helium which has the direct effect of increasing the SWCX X-ray emission.

By scanning the sky through the focusing cone, the spatial signature of the SWCX can be identified, allowing a separation and subsequent investigation of LHB and SWCX emission, such as the distribution of the hot plasma within the LHB. Combining this information with the geometry of the local cavity derived from other wavelengths it is then possible to derive the physical parameters of the plasma (its pressure and density). Knowing the physical conditions of the plasma will lead to more accurate pictures of the solar neighborhood and the evolution of bubbles of hot gas produced by supernovae or stellar wind when they near the end of their existence. Identifying and characterizing the SWCX signature will also permit an estimate of the contribution of heliospheric SWCX to the University of Wisconsin (McCammon et al. 1983) and *ROSAT* all-sky maps (Snowden et al. 1997) of the soft x-ray background, and to every current and future investigation of extended sources.

# 2. The Science Case for DXL

## 2.1. Diffuse X-ray Emission from the Local Galaxy

The soft X-ray diffuse background observed at ¼ keV with ROSAT (Snowden et al., 1997) originates from three main sources. A portion of it is extragalactic in origin and is due to the unresolved emission from discrete objects, typically active galactic nuclei (AGN), and the emission from hot filaments of the cosmic web (e.g., Hasinger et al. 1998, Mushotzky et al. 2000, Giacconi et al. 2001, Mateos et al. 2008, Gupta & Galeazzi 2009a, Nicastro et al. 2005, Ursino et al. 2006, Werner et al. 2008, Galeazzi et al. 2009, Fang et al., 2010). A second portion arises as thermal emission from hot gas in the halo of the Milky Way, probably at relatively low scale heights (<1 kpc – e.g., Galeazzi et al. 2007, Gupta et al. 2009b). A third portion of the emission must originate locally as X-rays are detected from the direction of the Galactic disk, where they would be easily absorbed by the neutral hydrogen in the Galaxy. One optical depth at ¼ keV is roughly $1 \times 10^{20}$ HI cm$^{-2}$, a column density, in the Galactic plane, that is reached in most directions within 100 pc.

Starting in the late 1970's, investigators suggested that the observed ¼ keV thermal emission originates in an irregularly shaped cavity in the neutral hydrogen of the Galactic disk that contains the Sun (Knapp 1975, Sanders et al. 1977) dubbed the Local Hot Bubble (Cox and Snowden 1986, Snowden et al. 1990). From the Wisconsin rocket data and later from the *ROSAT* All-Sky Survey (RASS) data, they argued that the emission is due to a thermal plasma with $T \sim 10^6$ *K*.

In the 1990's the Diffuse X-ray Spectrometer (*DXS*) was specifically designed to study the LHB by looking at the X-ray emission in the ¼ keV band with high energy resolution Bragg crystals. *DXS* performed a scan near the galactic plane and detected the presence of emission lines in the spectrum of the "local" diffuse emission, "confirming" the thermal nature of the source. However, exhaustive efforts to fit the spectrum globally with existing emission models, including non-equilibrium effects, multiple temperatures and variable abundances, were notably unsuccessful (Sanders et al 2001). The analysis of *DXS* data was performed assuming that all the emission originates in the LHB. The subsequent discovery of SWCX emission and the current inability to separate it from LHB emission, however, requires a reevaluation of the scientific results.

Also in the 1990's, the *ROSAT* All-Sky Survey (RASS) arguably provided the best maps of the ¼ keV and ¾ keV backgrounds. The angular structure of the emission was consistent with previous maps of the diffuse background but with superior statistics. However, during the course of the six-month survey it became apparent that there was a significant time-variable background that was spectrally indistinguishable from the low energy diffuse X-ray emission. There was no explanation for the physical origin of this background at the time so it was treated empirically and subtracted from the data, but it was understood that this contamination, referred to as "long-term enhancements" (LTEs), could not be entirely removed. As a result, all subsequent studies of the local component of the soft X-ray background using *ROSAT* data and, by extension all other observations, are plagued by the uncertainty of an unknown zero-level offset (consistent between surveys).

Independently, observations of comets with *ROSAT* showed them to be unexpectedly strong sources of soft X-rays (Lisse 1996). In the late 1990's SWCX, due to solar wind interaction with cometary neutrals, was determined to be the emission mechanism (Cravens 1997). Also in the late 1990's, Cox (1998) suggested that the solar wind should also charge exchange with interstellar neutrals passing through the solar system, providing a relatively constant diffuse emission in the soft X-ray band (E<1 keV).

In the 2000's, Cravens, Robertson and Snowden (2001) and Robertson and Cravens (2003a,b) suggested that the solar wind could also charge exchange with exospheric material in Earth's magnetosheath, providing a similar spectral background but with considerably more temporal variability. The magnetosheath emission, because of its temporal variability, is the origin of the LTEs observed by *ROSAT*. It is also likely to be the origin of some of the time variable contamination observed by the Wisconsin (McCammon et al. 1983), SAS-3 (Marshall and Clark 1984), and HEAO-1 A2 (Garmire et al. 1992) ¼ keV surveys.

In the 2000's it was suggested that SWCX emission may be responsible for all of the "local flux" observed at ¼ keV (Lallement 2004a,b), challenging the "conventional" interpretation of the low-energy DXB origin. There is now a controversy as to whether roughly half of the observed ¼ keV background originates within the nearest 100 AU or from a region that extends to several hundred parsecs, a dynamic range, and therefore uncertainty, of five orders of magnitude in the location of the source. The controversy about the location of "local" emission is not limited to the ¼ keV band. It has also been suggested that much of the observed ¾ keV emission may also originate from SWCX (Koutroumpa et al. 2007), affecting our understanding of the physical parameters of the intragroup medium and the halo of the Milky Way.

**2.2 Contamination of Extended Objects with Current Missions**

Contamination by SWCX impacts the accuracy of observations of low surface brightness objects that fill the field of view of X-ray observatories, such as the soft X-ray background, hot plasmas in nearby superbubbles, the interstellar medium in our galaxy, the Magellenic clouds, nearby galaxies, and even clusters of galaxies. Current analysis procedures remove time intervals showing variation, however, this does not identify the slowly changing heliospheric SWCX emission and results in a misleading interpretation of the spectral content of the source (e.g., Henley 2008). It is essential to construct effective models of the foreground SWCX emission and validate these models experimentally, otherwise the absolute calibration of all major X-ray observatories for soft, low-surface brightness objects will be uncertain and spectral modeling will lead to incorrect physical interpretation of sources. For example, Snowden et al. (2004) have looked at the temporal history of a variety of parameters for a 2001 *XMM-Newton* observation of the Hubble Deep Field North (HDFN). They observed strong SWCX emission with lines from C VI, O VII, O VIII, Ne IX, and Mg XI.

An accurate calibration of the SWCX emission is necessary not just for current missions (*XMM-Newton*, *Chandra*, and *Suzaku*) but of course for future missions as well, such as *eROSITA*, *Astro-H*, *IXO*, or *Athena* (Predehl et al. 2010, Takahashi et al. 2010, Bookbinder 2010). The *eROSITA* mission will be primarily an all-sky survey which will have good sensitivity in the ¼ keV and ¾ keV bands. An accurate knowledge of the SWCX background will be vital to studies of the diffuse X-ray background and for producing maps of the diffuse background alone. The other missions will be operated almost exclusively in pointing mode. With typically smaller fields-of-view than current missions, the difficulty of determining an off-source background will be increased and for many objects, e.g., nearby clusters, will be impossible. A model for the SWCX emission based on observation geometry and solar wind parameters will be critical for the proper interpretation of the data.

Except for the temporal signature due to changes in the solar wind flux, abundances, and ionization states, the SWCX background can be indistinguishable from low redshift astrophysical plasmas. Line emission from many of the same ions dominates both. Above 0.3 keV these are mostly K-shell lines of C, N, O, Ne, and Mg, and L-shell lines of Fe. Below 0.3 keV these are typically L- and M-shell emission from metals. Above 0.3 keV many of the emission lines (and those of O, in particular) are useful for identifying the temperatures, density, and ionization equilibrium of the diffuse thermal plasmas typically found in supernova remnants, superbubbles, and other interstellar plasmas, as well as emission from groups and clusters of galaxies. Many of these emission regions, both in the Milky Way and in nearby galaxies (e.g., the Magellanic Clouds) subtend large solid angles relative to the current instrumentation so observations cannot be "cleaned" of this contamination using data from the observation itself. Using a separate pointing to subtract the background is also unfeasible. As a result, without a clear understanding of the SWCX background there will be complete uncertainty in interpreting some observations and there will always be some level of uncertainty in the interpretation of all observations of astrophysical plasmas and this will be the case for all observatories.

**2.3 SWCX with Current and Future Missions**

The critical parameter for diffuse emission investigations is the *grasp* of the instrument, defined as the product of effective-area times field-of-view. Current missions are ineffective at detecting the spatial signature of the SWCX due to their limited grasp. A study of the spatial signature of the SWCX, such as the one carried out by DXL, with current satellites requires a large number of individual pointings and, to accumulate enough statistics, each pointing requires a significant amount of observing time. As an example, the DXL grasp at 0.65 keV, the O VIII energy, is about 7.5 cm$^2$ sr (600 cm$^2$ area, 6.5$^o$ x 6.5$^o$ FOV - see Fig. 1). By comparison, XMM-Newton has a grasp about 150 times smaller and Suzaku about 2,000 times smaller, requiring 40,000 s and 600,000 s of observing time respectively, to obtain at 0.65 keV the same science results as 5 minutes with DXL. At lower energy the situation is much worse, as in the ¼ keV band their effective area drops well below 100 cm$^2$, making any SWCX science there essentially impossible. Future calorimetric missions such as Astro-H and Athena will fare even worse, due their small field of view, while *eROSITA* is expected to have a grasp about 50 times smaller than DXL and will be used primarily in survey mode.

Moreover, even assuming that the necessary observing time is available, the interpretation of the data is highly complicated by the temporal variation of the SWCX emission that affects the self consistency of each observation and makes a comparison between observations difficult. For example, Koutroumpa et al. (2009a) investigated the flux in O VII and O VIII during on-cloud and off-cloud observations of MBM12 with Suzaku (Smith 2007). The two observations were performed consecutively to minimize the temporal variation of the SWCX component, but the time necessary to accumulate significant statistics was longer than the time in which the SWCX varied by a factor of more than 2, affecting the self consistency of each observation and making a direct comparison of the two observations impossible.

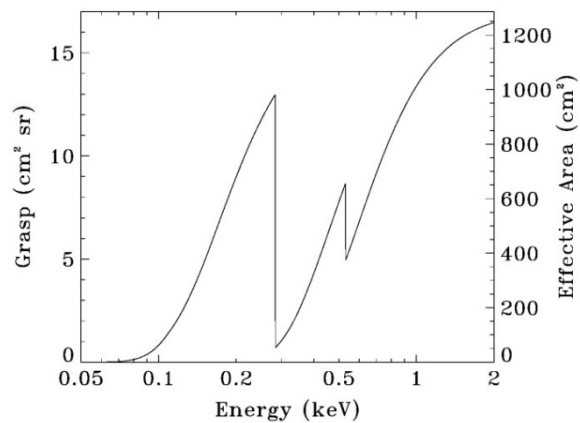

**Fig. 1.** Effective area and grasp of the DXL payload below 2 keV.

The X-ray Quantum Calorimeter (XQC – McCammon et al. 2002) sounding rocket program has the resolution in the soft band (<10 eV at 1 keV) to observe the emission from SWCX in the ¾ keV band and to study the spectral signature of the soft x-ray background, including the SWCX and LHB contributions with a high energy resolution microcalorimeter. However, XQC does not have the collecting area, angular resolution, or temporal scale to constrain global SWCX foreground models, and DXL, by verifying the spatial models of the SWCX emission, is a critical complement to the XQC results.

In the future, large collecting area, wide field of view instruments such as the proposed Lunar X-ray Observatory (Porter et al. 2008) and the Magnetospheric X-ray instrument (Collier et al. 2009) could provide global images on a moderate temporal time scale with moderate spectroscopic capability. However, at best, these experiments are 10-20 years in the future and will be more sensitive to SWCX due to the magnetosphere.

## 3. Mission concept

For the DXL mission we are refurbishing and will fly the existing University of Wisconsin (UW) Aerobee-IV payload. The Aerobee-IV payload (hereafter DXL) is composed of two large area (500 cm$^2$ each at 1 keV) thin window proportional counters that were used to produce the original UW all-sky soft X-ray maps. The large area of the proportional counters provides excellent counting statistics when scanning the sky through the He focusing cone in the limited observing time of a sounding rocket flight. For the scan we have chosen a region sufficiently close to the galactic plane to absorb soft X-rays produced outside the solar system and its local environment (within ~100 pc), therefore minimizing the intensity and structure of the cosmic background, as indicated by the RASS. The DXL payload is scheduled to launch from White Sands Missile Range in New Mexico around 2012 December to ensure a scan path through the He focusing cone when the sun is close to the maximum of its 11 year cycle. The time of the observation is important due to the dependence of the solar wind characteristics from the solar cycle. In fact, during solar maximum the high latitude SWCX emission should be higher, because of the higher heavy ion abundances, and the probability of eruptive events with higher solar wind fluxes and/or abundances is increased, which could result in higher SWCX emission. However, towards low latitudes (which is the DXL case) the quiescent SWCX should be lower during solar maximum. The SWCX emission will still be significant to see its spatial signature, but won't overshadow the LHB emission.

### 3.1. The He Focusing Cone

The bulk of the soft X-ray emission comes from SWCX with three neutral populations: interstellar hydrogen, interstellar helium, and the exospheric (geocoronal) hydrogen. SWCX due to geocoronal hydrogen may generate strong "bursts" of X-ray emission, which are clearly identifiable in X-ray observations. The flow of interstellar neutrals through the solar system is due to the motion of the heliosphere (at about 25 km s−1) through the local interstellar cloud (LIC). This material, a gas of mostly hydrogen atoms with about 15% helium, flows from the direction of l~252$^o$, b~9$^o$. This places the Earth upstream of the Sun in the interstellar neutral flow in early June and downstream in early December every year (Gruntman 1994). Interstellar hydrogen experiences gravitational force, an almost equal repulsive force due to radiation pressure, and significant ionization due mainly to charge exchange with high speed solar protons (e.g., Quémerais 1999, Lallement 1999), creating a hydrogen cavity around the sun. X-ray emission due to SWCX with interstellar hydrogen shows little geometric variation (dependence on location and look direction). However, only gravity significantly affects the helium trajectories which therefore execute Keplarian orbits and form a "focusing cone" downstream of the Sun centered at ~6$^o$ below the ecliptic plane, resulting in a localized downstream enhancement of helium observed annually by Earth orbiting and L1 spacecrafts (Bzowski 1996, Frisch 2000). The cone has a FWHM opening angle of about 25 degrees and at one AU its density peaks at ~0.075 cm$^{-3}$ compared to a more typical value of ~0.015 cm$^{-3}$. The peak density drops with distance from the Sun with a value of ~0.05 at 5 AU.

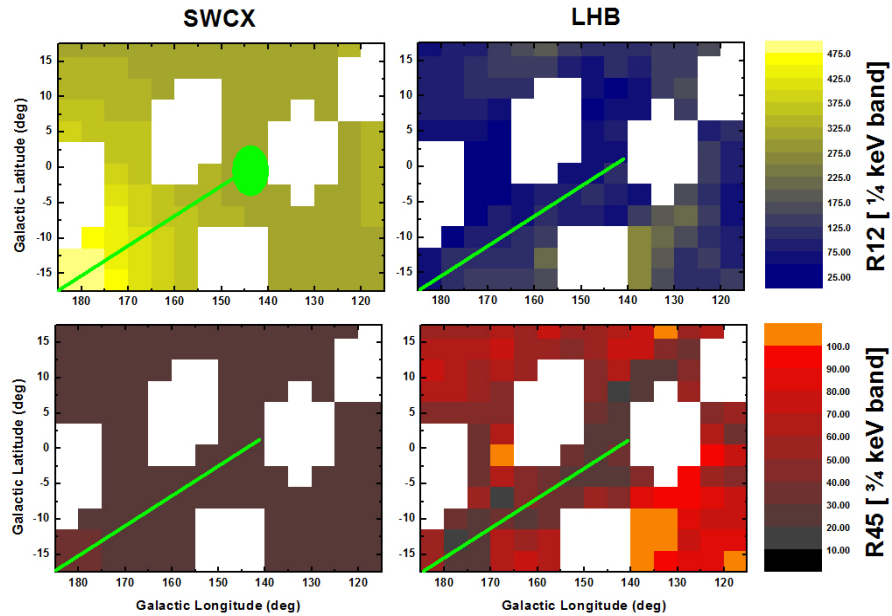

**Fig. 2.** Expected contribution of LHB and SWCX in the ¼ keV (R12) and ¾ keV (R45) energy bands with the mission scan path (green line) and aperture (green oval). The units are $10^{-6}$ counts s$^{-1}$ arcmin$^{-2}$. The spatial variation due to the He focusing cone is particularly evident in the ¼ keV band, where the SWCX emission varies by almost a factor of 2.

While there are well-defined models for SWCX emission from the heliosphere, testing them is problematic. Koutroumpa et al. (2007), using a self-consistent model of the heliospheric SWCX emission, managed to associate observed discrepancies in XMM-Newton and Suzaku observations separated by several years with solar cycle-scale variations of the heliospheric SWCX emission. A recent paper (Snowden 2009) was successful in using an XMM-Newton observation to test a model (Robertson & Cravens 2003a) for emission from the near-Earth environment. Koutroumpa et al. (2009b) uses a different series of XMM-Newton observations to search for SWCX emission from the He focusing cone. They were relatively successful, but the detection was at a marginal level due to the short exposures and higher backgrounds experienced by XMM-Newton.

**3.2. Observing Strategy**

The last solar maximum, solar cycle 23, occurred in 2001. The DXL payload will be launched from White Sands Missile Range in New Mexico during the month of December 2012 allowing a scan path through the He focusing cone when the sun is close to the maximum of its 11 year cycle. Contrary to what may be expected, a solar maximum corresponds to a lower SWCX emission because the higher solar flux ionizes the bulk of the H in the inner solar system, removing many of the neutrals involved in the charge exchange. The SWCX emission will still be sufficiently strong to see its spatial signature, but won't overwhelm the LHB emission.

Summarizing what has already been mentioned, the observing strategy is based on the following guidelines:

- Launch in early December when the He focusing cone is aligned with the Sun-Earth line, directly opposite the Sun;
- Limit the scan to ±20° from the galactic plane to assure that only "local" emission is detected (<100 pc). This will remove any contribution from DXB components other than LHB and SWCX;

- The low galactic latitude will also avoid bright high latitude emission regions visible in the RASS maps;
- Avoid significantly bright point sources;
- Scan a region of the sky where the LHB emission is expected to be relatively uniform.

For the first DXL investigation we have chosen a region of the sky roughly centered at l~150°, b~0°, shown in Fig. 2. The figure shows the expected contribution from the LHB and the SWCX in the ¼ and ¾ keV bands. The LHB contribution is calculated using RASS maps from which we subtracted our model of the SWCX for the RASS observing period and geometry, the SWCX contribution is calculated using our model for DXL geometry at solar maximum (Koutroumpa et al. 2006, 2007). Notice that, due to the different observing geometry, the RASS observations of this region were not significantly affected by the He focusing cone enhancement. The empty pixels correspond to regions with bright objects. The effect of the He focusing cone is clearly evident in the lower left corner of the ¼ keV SWCX map and corresponds to an enhancement of about a factor of 2 in the SWCX emission. The green lines in the figures correspond to the chosen scan path. The observation will start roughly in the direction of the He focusing cone, at l=185°, b=−18°, move to l=140°, b=0° and back. During the observation, in the ¼ keV band the SWCX emission is expected to be the biggest component and change from about $300\times10^{-6}$ counts s$^{-1}$ arcmin$^{-2}$ (RASS Units≡RU from now on) to more than 500 RU, while the remaining DXB emission should be roughly constant at 100 RU. In the ¾ keV band the SWCX emission should change from about 20 RU to about 30 RU, while the remaining DXB should be roughly constant at about 30 RU. The combination of spatial information in the two separate bands should allow a clear separation and characterization of both components. The DXL count rate in both bands as a function of galactic longitude is shown in Fig. 3. Assuming resolution elements of 6°×6°, equal to the DXL resolution, a 5 minute flight will provide about 40 s observing time per resolution element, or between 6,000 and 10,000 counts per resolution element in the ¼ keV band and about 800 counts per resolution element in the ¾ keV band, providing very high statistics for the data analysis. We note that we do not expect our count rate to be affected by any bright point source which will stay clear of our aperture. We also expect to have post-flight pointing accuracy of about 1 degree, which would allow us to use RASS to quantify and remove any such contribution in case the actual trajectory is significantly off the expected one.

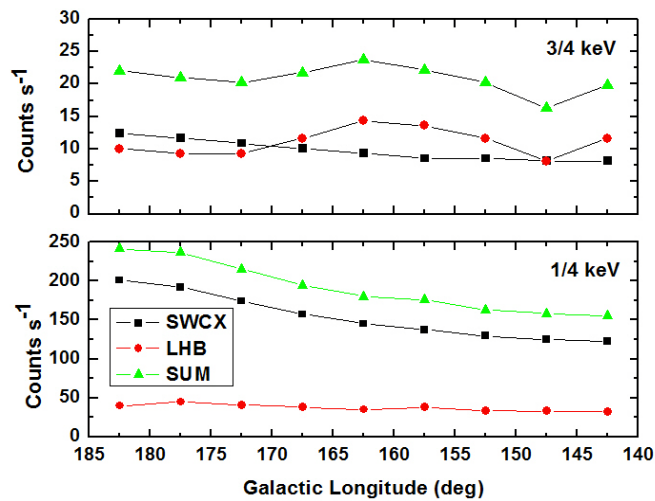

**Fig. 3.** Expected DXL count rates in the ¼ keV and ¾ keV energy bands across the DXL scanning path due to LHB and SWCX emission.

### 3.3. Data Analysis and Expected Results

The signal that DXL will measure is the sum of the SWCX emission due to the solar wind interactions with He and H, and the cosmic background emission. By observing in the downstream direction of the interstellar neutral flow we minimize both the total contribution and the spatial variation of the H SWCX. The total He SWCX intensity $I$ is the integral over the line of sight $ds$ of the product of the solar wind density $n_{SW}$, the He density $n_{He}$, and the compound interaction cross-section $\alpha$

$$I = \int \alpha n_{SW} n_{He} <g> ds ,$$

where $<g>$ is the averaged total solar wind speed (both bulk and thermal). The compound cross-section is the abundance-weighted sum of the cross-sections of all the different species and ionization states in the solar wind. The individual cross-sections are velocity dependent and are poorly known, even for some of the dominant species. The Advanced Composition Explorer (ACE) satellite data (Stone et al. 1998) provides many of the important solar wind abundances and ionization state fractions for the solar wind as it passes the Earth. Combining the ACE data with a model of the solar wind we will reconstruct the solar wind density, velocity, and, more roughly, the ionization structure along the line of sight as a function of time. The shape of the profile in Figure 3 is determined by the solar wind parameters and the distribution of the neutral He, both of which are reasonably well understood. The amplitude of the profile is instead determined by $\alpha$, which is not well understood.

As first step, by fitting our observed profile with the solar wind and He models we can determine $\alpha$. By simultaneous fitting of DXL data and the ROSAT all-sky survey data (together with the IMP-8 solar wind data – http://science.nasa.gov/missions/imp-8/) we can constrain the free parameters of the solar wind model, such as the latitude extent of the equatorial component of the solar wind. We then use the models and the newly determined values of $\alpha$ to remove the He SWCX from the RASS. With a better understanding of the SWCX parameters we will also repeat the several historic analyses that constrained the nature of the LHB to determine its properties if, in fact, there is still emission not accounted for by SWCX.

We note that, in our model, we have additional spectral information, that treats individually each ionization state, although in the ¼ keV the uncertainties on the cross-sections may be large, as well as the various line emission probabilities. We will therefore also be working on the shape of the spectrum in the 1/4 keV, which is critical to determine the emission mechanism.

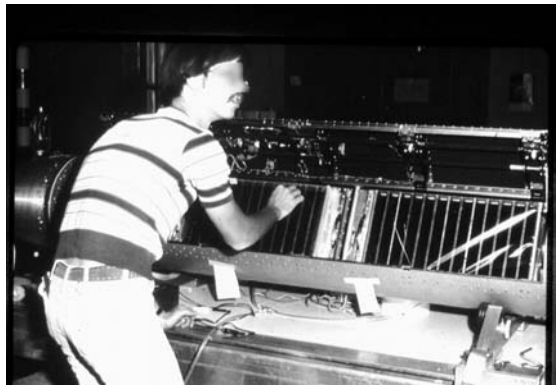

**Fig. 4.** DXL re-uses the University of Wisconsin Aerobee-IV payload shown here. The payload consists of two thin windowed proportional counters with a total effective area of about 1000 cm$^2$ at 1 keV.

# 4. The Instrument

The Wisconsin All Sky Survey program was based on a set of ten sounding rocket flights with mechanically collimated proportional counters that flew between 1972 and 1980 (McCammon 1983). Two different generations of the experiment were used in the survey, the Aerobee-III payload (1972-1973) and the Aerobee-IV payload (1973-1980). For the DXL payload we are refurbishing the newer, larger, Aerobee-IV payload that is shown in Figs. 4 and 5.

The DXL payload consists of two thin-windowed, large area proportional counters mounted on a magnesium frame and supported by the rocket skin. The counters use a wire-wall design (multi-anode proportional counters with wire cathode walls) and are filled with P10 gas (90% Argon, 10% methane) at 760 torr. The X-rays are mechanically collimated using 2.5 cm thick, 3 mm cell honeycomb providing a circular field of view of 6.5° FWHM. Ceramic magnets embedded in the collimator provide a magnetic field of 150 Gauss within and above the collimators which intercepts electrons that could either generate scattered X-rays or mimic X-ray events in the detectors. A three-sided veto for cosmic-ray rejection is also obtained using anodes at the sides and back of each main counter.

One side of each counter, facing the mechanical collimators, is covered with a thin X-ray window (25 cm x 50 cm). Different window materials have been used on the Aerobee-IV experiment to provide different energy responses (Williamson and Maxson, 1975). For this investigation we will use thin (80-90 mg cm$^{-2}$) plastic windows supported by a 100 lines-per-inch nickel mesh (with ~68% transmission) to retain the counter gas while allowing soft X-ray photons to enter the counter. The windows are composed of Formvar ($C_5H_7O_2$, the registered trade name of the poly-vinyl formal resin produced by Monsanto Chemical Company) with an additive, Cyasorb UV-24 ($C_{14}H_{12}O_4$, a trade name of American Cyanamide), to absorb stellar ultraviolet photons that could otherwise generate a large non-X-ray background. The same windows were used for the DXS experiment (Sanders et al. 2001) that flew on-board the Space Shuttle *Endeavour* in January 1993. The use of the DXS windows, rather than the Wisconsin all-sky survey ones, increases the effective area at O VII by about a factor of three, increasing the DXL ¾ keV band sensitivity. Geometrical effects, including the shadow of the nickel mesh, anticoincidence detectors, collimators, magnets, and masks near the ends of the anode wires reduce the effective area of each detector to about 600 cm$^2$. The effective area as a function of energy, including the window transmission, is shown in Fig. 1 for the sum of both counters. While the energy resolution of the proportional counters is relatively poor below 1 keV, the carbon edge at 0.285 keV in the filter response allows a clear separation of the events below 1 keV in two different bands, corresponding roughly to the *ROSAT* 1/4 keV and 3/4 keV bands.

To ensure gain stability, a constant gas density is maintained inside the counters during the flight by using an on-board gas bottles and pressure regulators. The regulators are true density regulators, using differential pressure between the main counter and a reference volume at the same temperature. A high-stability, high-voltage power supply provides 1,700V bias for the counters. Two $^{55}$Fe calibration sources (one for each counter) with mechanical shutters allow gain calibration during the flight.

The signal from the proportional counters is read out using low noise charge amplifiers digitized, and telemetered to the ground. X-ray pulse heights are telemetered with a resolution of about 2 eV from 30 eV to 1 keV and 20 eV from 1 keV to 10 keV. The energy resolution is therefore limited by the intrinsic resolution of the proportional counters and is equal to about $15 \times \sqrt{E[eV]}$ FWHM. A second set of multiple threshold discriminators also provides an estimate of the pulse amplitude as a backup for the main pulse acquisition system.

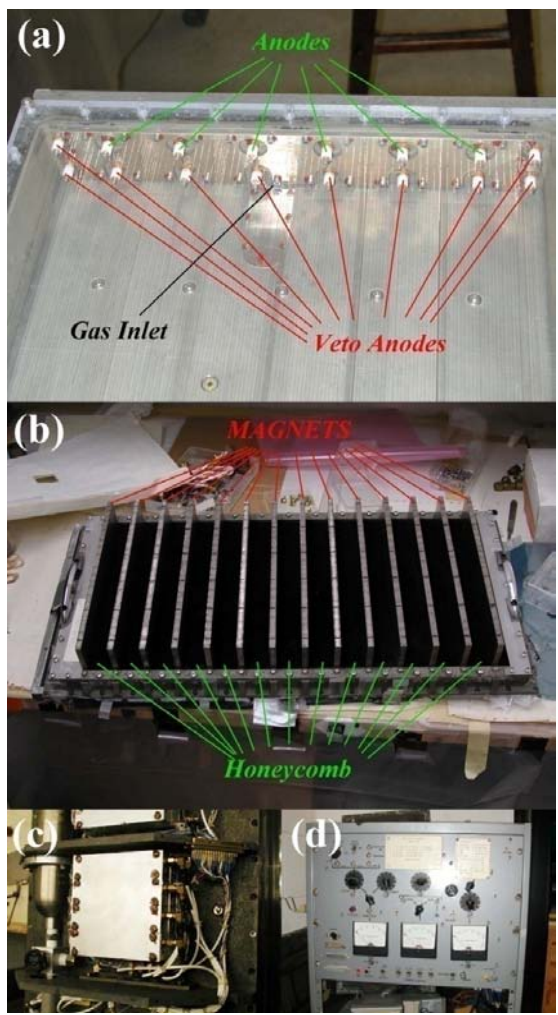

**Fig. 5.** DXL hardware: (a) proportional counters, (b) collimator and magnets, (c) electronics, and (d) GSE.

# 5. Beyond the first flight

The DXL mission has, so far, been approved by NASA for the payload refurbishment and the first flight described in this paper. However, we foresee this as a longer term campaign for a complete investigation of the properties and characteristics of the LHB and SWCX, with additional flights beyond 2012.

As noted before, our SWCX model, and in general any proper interpretation of SWCX emission, require additional spectral information by treating each ionization state individually. However, in the ¼ keV the uncertainties on the cross-sections, as well as the various line emission probabilities, may be large. The XQC payload will be invaluable in that respect, but will lack the spatial resolution and grasp to fully separate SWCX from LHB emission and provide a high statistics measurement. The proportional counters on DXL do not have the energy resolution necessary to separate the lines generated by the various ionization state. However, additional energy information can be obtained by using different "filters" in front of the counters. For example, as it was done with the Wisconsin all sky survey, it is possible to replace the (mostly carbon) window in front of the counter with either a boron or a beryllium window to investigate different energy bands in the ¼ keV range (see, for example, McCammon & Sanders 1990, Figure 3). The refurbished DXL payload is being designed with the necessary infrastructure to accommodate, in future flight(s), a smaller, 150 cm$^2$ counter capable of accommodating either a Be or B window, based on the counter design used by Bloch et al. (1986). It will then be possible to use detailed data from ACE for each ion species to separate the different ion contributions. This has been unnecessary until now, since XMM, Suzaku and Chandra primarily allow only the oxygen (and perhaps C, N and Ne) lines to be resolved.

Moreover, although heliospheric SWCX is the primary concern for future X-ray missions, a good understanding of SWCX mechanisms requires also a study of its magnetosheath component. As mentioned before, magnetosheath SWCX is, in general, small with respect to the heliospheric component, but may become the dominant term during periods of high solar wind flux. The X-ray emission due to magnetosheath SWCX is also affected by a significant spatial variation, in this case due to the solar wind direction with respect to the Earth motion (see, for example, Porter et al. 2008). Such a spatial variation can be used, similarly to what has been described in this paper, to separate magnetosheath SWCX from other X-ray contributions to the DXB. However, the emission is significantly affected by the solar wind strength and composition, and the investigation requires a launch synchronized with solar wind activity. Solar wind data are provided by the ACE satellite (Stone et al. 1998) and available, in almost real time, through NOAA (http://www.swpc.noaa.gov/SWN/index.html). As solar wind activity cannot be predicted in advance, the investigation requires a continuous monitoring of ACE data while waiting for the proper solar wind conditions. Such a wait may last for days and it is not feasible from White Sands Missile Range due to the high cost and limited availability of launch dates. However, "the Sounding Rocket Program is planning to provide launches from Woomera, Australia, in the fall of calendar year 2014 and again in the spring of calendar year 2016, subject to the availability of funds" (ROSES-10 Amendment 26). Launching outside a military facility, with significantly smaller requirements would be the perfect opportunity for such a triggered observation. A natural follow-up to the first DXL flights could therefore be two additional triggered flights from Woomera, one during a period of high solar wind flux, the during a "average" period.

**Acknowledgments** The investigation is supported by the National Aeronautics and Space Administration (NASA), grant # NNX11AF04G. DK acknowledges support from a NASA Postdoctoral Program fellowship administered by Oak Ridge Associated Universities.